\documentstyle[epsf, rotate]{article}
\begin{document}
\begin{center}
{
\Large {\bf On the nature of the compact X-ray source inside RCW 103}

{\bf Sergei B.Popov }

Sternberg Astronomical Institute,
 Universitetskii pr.13, 119899, Moscow, Russia\\
e-mail: polar@xray.sai.msu.su

}
\end{center}

\begin{abstract}
 I discuss the nature of the compact X-ray source inside the
supernova remnant RCW 103. Several models, based on the accretion
onto a compact object such as a neutron star or a black hole
(isolated or binary), are analyzed. 
I show that it is more likely that the X-ray source is an
accreting neutron star than an accreting black hole. 
I also argue that models of a binary system with an old accreting neutron star 
are most favored.

\end{abstract}

\section{Introduction}

It is generally accepted that most of  neutron stars (NSs)
and black holes (BHs) are the products
of  supernova (SN) explosions (although there is also a possibility of a
``quiet collapse''). In some cases a supernova remnant (SNR)
appears after a {\it formidable} explosion of a massive star.
Although a young NS is often observed inside a SNR 
as a radio pulsar (e.g., Crab, Vela, etc.), 
in most cases no compact object is found inside a SNR, or an
accidential coincidence
of the radio pulsar and the SNR is very likely (e.g., 
Kaspi 1996; Frail 1997).

Recently, Gotthelf et al. (1997) found a compact X-ray source 
inside the  SNR RCW 103 with the 
X-ray luminosity $L_x\sim 10^{34}\, erg/s$ (for the distance $3.3\, kpc$)
 and the black-body temperature  about $0.6\, keV$.
 The nature of the compact source is unclear. No radio or optical compact 
counterpart was observed. Here, I discuss possible models of that
compact source.

\section{What is inside the RCW 103?}

Gotthelf et al. (1997) discussed why the source cannot be
a cooling NS, a plerion, or a binary with a normal companion.
The reader is referred to their paper for the details.
In the present analysis I assume that the X-ray luminosity of the 
source is  produced due to accretion of the surrounding material 
onto a compact object (a NS or a  BH). 
I analyse thus only models with compact objects, isolated or with compact
companions. Massive normal companions are excluded by optical observations.
If the companion is a low-mass star, it is difficult to
explain the X-ray luminosity as high as observed in the RCW 103
because in low-mass systems accretion usually occurs after the 
Roche lobe is overflowed with higher luminosities. 

 The main challenge for the models of accretion of the surrounding material
is to answer the question of
where the compact object finds enough matter to accrete. 
I don't discuss it here, assuming that
the material is available in the surrounding medium.

\subsection{Accreting isolated young black hole or accreting
old black hole in pair with a young compact object}

An isolated BH accreting the interstellar medium can be, in principle, observed
by X-ray satellities such as $ROSAT$, $ASCA$ etc (Heckler \& Kolb 1996).
Using the fact that neither radio pulsar nor X-ray pulsations are 
actually observed in the case of RCW 103, one can
argue that after the SN explosion a BH is formed more likely than 
a NS.
We then can explain why a compact X-ray
source inside a SNR without a radio pulsar are rare: the BHs are born
by the most massive stars, and so BHs  are an order of magnitude
less abundant than the NSs. 

To achieve high X-ray luminosity, 
a compact object must move with a relatively low velocity (the 
Bondi's formula):

\begin{equation}
 \dot M=2 \pi \left(
    \frac{(GM)^2 \rho}{(V_s^2+V^2)^{3/2}}\right),
\end{equation}
where $V_s$ is the speed of sound, $V$ is the velocity of the
compact object with respect to the ambient medium, $M$-- the mass of the 
accreting star and 
$\rho$ is the density of the accreting material.
One can introduce the effective velocity, $V_{eff}$, and rewrite eq. (1)
 as follows:
$$
\dot M=2 \pi \left(
    \frac{(GM)^2 \rho}{V_{eff}^3}\right).
$$

The effective velocity cannot be much lower than $10\, km/s$,
which corresponds to the sound speed in the ISM with the temperature 
of $\sim 10^4\, K$.
I will therefore use this value of the velocity, $10 \, km/s$, because
the luminosity is high for an isolated object, and with the lower velocity
much higher density of the surrounding medium is required.

During the SN explosions a compact object can obtain an additional 
kick velocity. The value and the distribution of the kick velocity is 
not known well enough (e.g., Lipunov et al. 1996).
Although observations of radio pulsars favour high kick velocities 
about $300-\, 500 \, km/s$
(Lyne \& Lorimer 1994), alternative scenarios in which the velocity
increases after the SN explodes are possible (Kaspi 1996;
Frail 1997). The most popular scenarios usually predict the mean 
kick velocities to be much higher than $10\, km/s$.

To explain the observed X-ray luminosity of the compact object inside  RCW 103 
the accretion rate, $\dot M$, should be about $10^{14}\, g/s$
 (assuming that one gramm of accreted material produces
$10^{20} erg$).
For all models that consider accretion onto an isolated compact object,
the density required to obtain $L_x\sim 10^{34} \, erg/s$ is as high 
as $10^{-22}\, g/cm^3$.

One can then estimate the size of the emmiting region, 
using observed luminosity and temperature:

$$
 L=4\pi \cdot R_{emm}^2 \sigma T^4
$$

For observed values 
of $ L_x$ and $ T$ 
this equation gives $R_{emm} \sim 0.9 \,km$.
For BHs such a low value of $R_{emm}$ is very unlikely because
the gravitational radius is about $R_G\sim 3\,km\, (M/M_{\odot})$.
Also, the efficiency of spherically symmetric accretion onto a
BH is very low resulting in a significantly higher density required
to achieve the same luminosity.
This is probably the main argument against isolated accreting
black hole as a model for the RCW 103. 

The same argument can be used against models with a binary system: BH+BH
(BH is born in the recent SN explosion) and BH+NS (NS is born
in the recent SN explosion and the pulsar is not observed due, for example,
to unfortunate orientation), or against models in which no compact 
remnant survives after the recent SN explosion of the massive star
in a binary system with a BH as a companion. 

In the next subsections I present more viable models with accreting
NS.

\subsection{Accreting isolated young neutron star}

 In the past few years isolated accreting NSs have become a subject 
of great interest due to the
new observations by the $ROSAT$ satellite (Treves \& Colpi, 1991;
Walter et al. 1996; Haberl et al. 1996). 
 In this subsection I will present an argument that the compact X-ray
source in RCW 103 can be an isolated accreting NS
and will estimate some properties of that NS.

 There are four main possible stages for a NS in a low-density plasma:
$1).$ Ejector (or a radio pulsar); $2).$ Propeller; 
 $3).$ Accretor; and $4).$ Georotator
(Lipunov \& Popov 1995; Konenkov \& Popov 1997).
 The stage is determined by the accretion rate, $\dot M$, the magnetic field
of the NS, $B$, and by the spin period of the NS, $p$.

 If the NS is in the Accretor stage, then its period is longer than the
accretor period, $P_A$:

\begin{equation}
 P_A=2^{5/14}\pi \, (GM)^{-5/7} (\mu ^2/\dot M)^{3/7}\, sec,
\end{equation} 
where $\mu = B\cdot R_{NS}^3$ is magnetic moment of the NS.

For the RCW 103 I use the following values: $\dot M= 10^{14}\, g/s$,
$M=1.4\, M_{\odot}$, $R_{NS}=10^6\, cm$ which give:

\begin{equation}
B\sim 10^{10}\cdot p^{7/6}\, G.
\end{equation}

This critical line for this period (eq. 2), $P_A$, is shown in the Figure 1.
The region below the line is allowed for accreting NS.


If material is accreted from the turbulent interstellar medium, a new
equilibrium period can occur (Konenkov \& Popov 1997):

\begin{equation}
	P_{eq}\sim 30\, B_{12}^{2/3}I_{45}^{1/3}\dot M_{14}^{-2/3}
R_{{NS}_6}^2 V_{{eff}_6}^{7/3}V_{{t}_6}^{-2/3}M_{1.4}^{-4/3}\, sec,
\end{equation}
where $V_t$ is the turbulent velocity (all velocities are in units 
of $10\, km/s$); $M_{1.4}$ is the  mass of the NS in units of $1.4\, M_{\odot}$,
$B_{12}$ is the magnetic field of the NS in unites
$10^{12}\, G$ and $R_{NS}$ is the radius of the NS in units of $10^6\, cm$.

We then obtain:

\begin{equation}
  B\sim 8\cdot 10^{9}\cdot p^{3/2}\, G.
\end{equation}

The corresponding critical line for the equilibrium period (eq. 4)
is also shown in the Figure 1.

It is obvious from the Figure 1 that to explain the luminosity 
of the RCW 103 by an isolated accreting NS,
one must assume that the NS was born with extremely low magnetic 
field or with
unusually long spin period. The age of the SNR RCW 103 is about 1000 years
(Gotthelf et al., 1997), which
means that magnetic field could not decay significantly
(Konenkov \& Popov 1997). Thus, the model with 
isolated young accreting NS is not a likely explanation for the data.

\subsection{Accreting old neutron star in pair 
with a young neutron star or a young black hole}

 Binary compact objects are natural products of binary evolution
(Lipunov et al. 1996). One can, therefore, discuss these scenarios
as a viable alternative.

 In the previous subsection I showed that accretion onto a young isolated
neutron star requires unusual initial parameters. However, 
there is a chance that we observe a binary system, where one component is an 
old neutron star and the other component (a NS or a BH) 
was formed in a recent SN explosion 
(or there was no remnant at all).

 In that case, the parameters 
determined by eqs.(3), (5) are not unusual:
old NS can have low magnetic fields and long periods. Due to the fact
that  Gotthelf et al. (1997) did not find any periodic change of the 
luminosity,
 one can argue that the field is too low to produce
the observable modulation (the accreting material is not
channeled to the polar caps: $B <10^6\, G$) or that the period is very long
($p>10^4\, sec$), which is contrary to what is expected 
(Lipunov \& Popov 1995). 

 The evolutionary scenario for such a system is clear enough (Lipunov et
al., 1996). One can easily calculate it using the ``Scenario Machine''
WWW-facility ({\bf http://xray.sai.msu.su/sciwork/scenario.html}; 
Na\-zin et al. 1996). For example,
two stars with masses $15 \, M_{\odot}$ and $14 \, M_{\odot}$ on the
main sequence with the initial separation $200 \, R_{\odot}$, $R_{\odot}$ --
the solar radius, after 14 Myr  (with two SN explosions
with low kick velocities: about $60 \, km/s$) 
end their evolution as a binary system NS+NS.
The second NS is 1 Myr younger. During 1 Myr the magnetic field can decrease
up to 1/100 of the initial value with a significant spin-down (Konenkov \&
Popov, 1997). The binary NS+NS is relatively wide: $ 20\, R_{\odot}$ with an 
orbital period $5.8^d$, so the orbital velocity is not high (the orbital
velocity of the accreting NS should be added to $V_{eff}$).

Young NS can be unobsereved as a radiopulsar due to several reasons.
The simplest one is the effect of orientation.

\section{Conclusions}

To conclude, I argued that the most likely model for the RCW 103 is
that of an accreting old NS in a binary system with a young
compact object born in the recent SN explosion that produced
the observed supernova remnant (or no remnant survived after the explosion).
Such systems a rare, but natural products
of the binary evolution. Scenarios with single compact objects or with
accreting BH are less probable. 
  
{\bf Acknowledgements}

I thank M.E. Prokhorov for helpful discussions,
A.V. Kravtsov for his comments on the text 
and the ``Scenario Machine'' group
for the WWW-version of the program.
The work was supported by the RFFI (95-02-6053),
the INTAS (93-3364)  and the ISSEP (a96-1896) grants.

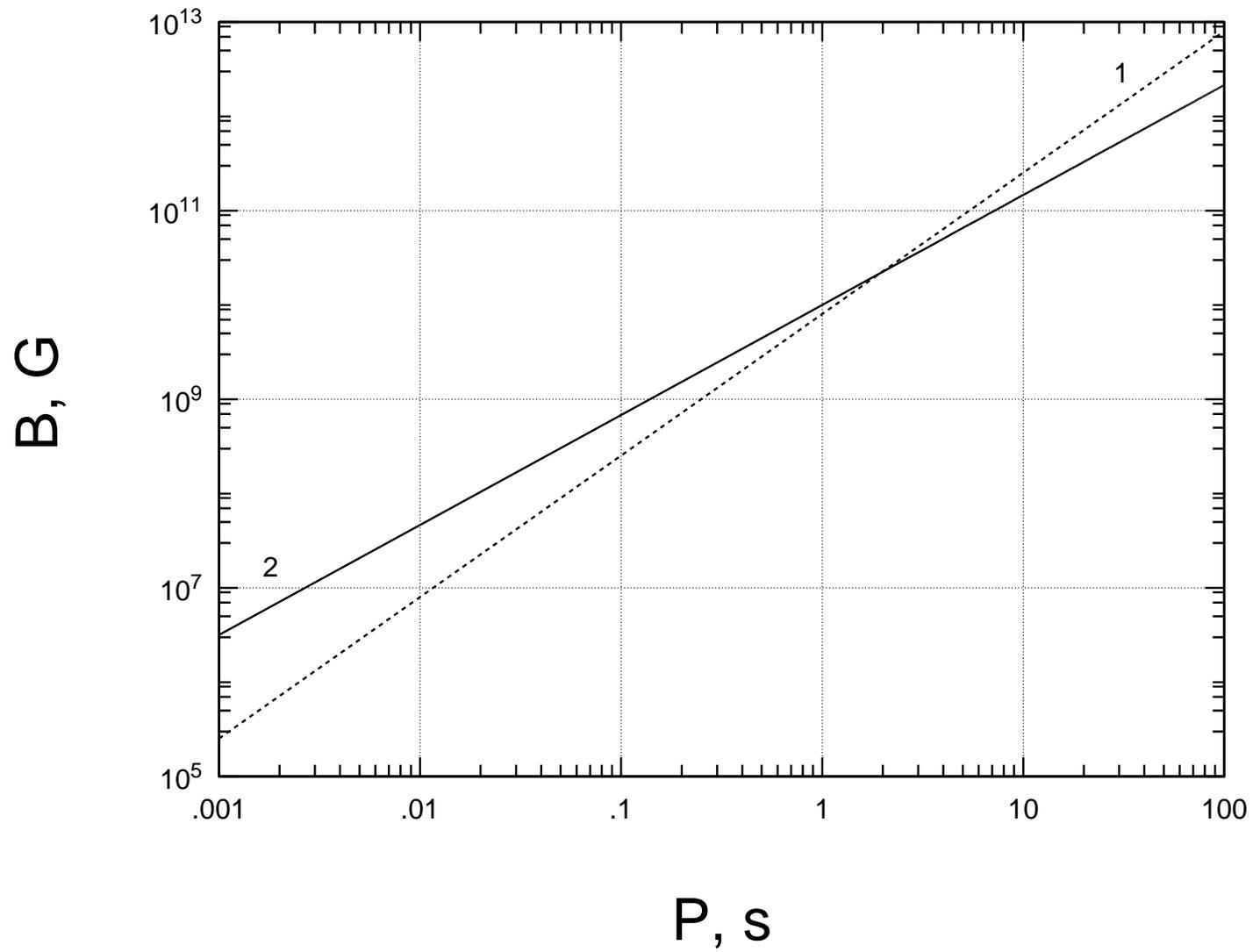
\end{document}